\font\titlefont=cmbx10 scaled\magstep1
\magnification=\magstep1
\mathsurround=2pt
\openup 2pt
\hfuzz 20pt
\def\a{A_{\mu\nu\rho}}
\null
\vskip 1.5cm
\centerline{{\titlefont The Dual Higgs Mechanism and the Origin of Mass in the
Universe}\footnote{$^*$}{{\it Honorable Mention} Awarded Essay by the
Gravity Research Foundation}}
\vskip 1.8cm
\centerline{{\bf A.Aurilia}\footnote{$^\dagger$}
{e-mail address: aaurilia@csupomona.edu} }
\smallskip
\centerline{Department of Physics}
\centerline{California State Polytechnic University}
\centerline{Pomona, CA 91768}
\bigskip\smallskip
\centerline{{\bf E. Spallucci}\footnote{$^\ddagger $}
{e-mail Address: spallucci@trieste.infn.it}}
\smallskip
\centerline{Dipartimento di Fisica Teorica, Universit\`a di Trieste, Italy}
\centerline{and}
\centerline{Istituto Nazionale di Fisica Nucleare, Sezione di Trieste, Italy}
\vskip 1.5cm
\centerline{\bf Abstract}
\smallskip\midinsert\narrower\narrower\noindent
The idea that a background of invisible material pervades the whole
universe is as old as the history of natural philosophy. Modern particle 
physics and cosmology support that idea and identify in the cosmic vacuum
the ultimate source of matter-energy, both seen and unseen, in the universe.
In the framework of the inflation-axion scenario, we suggest an unusual
realization of the Higgs mechanism which converts the latent energy of the
vacuum into observable mass. The existence of a new spontaneously broken
symmetry is pointed out which has the same effect of the Peccei-Quinn symmetry
in creating axions as pseudo-Goldstone bosons. However such particles are
eliminated from the physical spectrum in favor of massive pseudoscalar
particles.
\endinsert
\vfil\eject
The fundamental paradigms of modern physics are the standard Big Bang model
of cosmology and the standard 
$SU(3)_{\rm C}\otimes SU(2)_{\rm L}\otimes U(1)_{\rm Y}$ model of the strong
and electroweak interactions. During the past decade both models have been 
refined with the addition of two key ingredients: {\it inflation} on the
cosmological side and {\it axions} as pseudo Goldstone bosons associated with
the spontaneous breakdown of the Peccei-Quinn symmetry in particle physics.

Inflation requires the existence of dark matter and axions have long been
candidates for cold dark matter, even though there is no a priori reason
why the two notions should be related at all.

If the cosmological constant today is zero, and there are claims to the
contrary$^1$, the axion could be detected in an experiment capable of
probing masses in the range $10^{-6}\sim 10^{-3}$eV. Where does this mass
come from? The general consensus is that it comes from the quantum anomaly
which violates the chiral $U(1)_{\rm PQ}$ symmetry, thereby evading
Goldstone's theorem.

However, the chiral anomaly is just one of at least two possible loopholes
by which the existence of a Goldstone boson can be avoided . The second
loophole is the Higgs mechanism. Our objective is to suggest a variation
of the Higgs mechanism which appears to be tailor made for the axion-mass
problem but operates entirely within the framework of the inflationary
cosmology. We call this novel mechanism the `` dual '' (abelian) Higgs
mechanism, or DHM for short, since it applies to a totally antisymmetric 
tensor gauge field $\a$ which is the dual of the vector gauge field
encountered in the usual Higgs model.

In order to place DHM in the right perspective and partly to motivate it,
consider the inflationary idea that the early phase of the exponential
expansion of the universe inflated a microscopic volume of space 
to a size much larger
that the presently observable part of the universe; this idea can be formulated
within the framework of general relativity as a special case of `` Classical
Bubble Dynamics '' (CBD), i.e. the study of the evolution of a vacuum
bubble in the presence of gravity$^{2-4}$. In our own formulation of CBD$^4$,
inflation is driven by a temporary field $\a$ which is equivalent to a 
cosmological constant and DHM is the precise mechanism that converts the
self-energy of $\a$ into observable mass. How does this conversion take place?
The following properties of $\a$ constitute the crux of DHM in the
inflation-axion scenario:

a) First, $\a$ represents `` dark stuff '' by definition, since in 
(3+1)-dimensions $\a$ does not possess radiative degrees of freedom.
In fact, the field strength 
$ F_{\mu\nu\rho\sigma}\equiv\nabla_{[\,\mu}\, A_{\nu\rho\sigma\,]}=
\partial_{\,[\,\mu}A_{\nu\rho\sigma\, ]}$ is simply a constant disguised as a
gauge field. This property, even though peculiar, is not new in quantum
field theory: it is shared by all n-forms in n-spacetime dimensions. 
For instance in two dimensions, 
$F_{\mu\nu}=\partial_{[\,\mu}\, A_{\nu]}=\epsilon_{\mu\nu}\, f$ while in four
dimensions, $F_{\mu\nu\rho\sigma}=\epsilon_{\mu\nu\rho\sigma}\,f$, and $f$
represents a constant background field in both cases by virtue of the field
equations. What is then the meaning of ``$f$''? As a gauge field, $\a$ is
endowed with an energy momentum tensor and thus it couples to gravity: the
resulting equations are Einstein's equations with the 
cosmological term $\Lambda=4\pi\, G\,f^2$.
For this reason we call $\a$ the ``{\it cosmological field} ''.

b) Second, if the cosmological field acquires a mass, then it describes
massive {\it pseudoscalar} particles, in sharp contrast with the usual Higgs
mechanism. For instance, in Minkowski space the field equations
$$
\partial^\mu \, F_{\mu\nu\rho\sigma}-m^2\, A_{\nu\rho\sigma}=0
\eqno(1)
$$
are equivalent to the set of equations
$$
\left(\, \partial^2-m^2\, \right)\, A_{\mu\nu\rho}=0\ ;\qquad 
\partial^\mu\, A_{\mu\nu\rho}=0
\eqno(2)
$$
and the supplementary condition imposes three constraints on the four
components of $\a$ leaving only one propagating degree of freedom:
$\partial^\mu A^*_\mu$ where $A^*_\mu$ is the (pseudo)vector dual to
$\a$.

c) Third, $\a$ does not interact directly with ordinary matter fields.
Rather, it is the `` gauge partner '' of relativistic closed membranes,
or bubbles, in the sense that it mediates the interaction between surface
elements according to the same general principle of local gauge invariance
which dictates the coupling of point charges to vector gauge bosons. What
is then the gauge structure of bubble dynamics? It has long been known$^5$ that
relativistic extended objects possess a new kind of $U(1)$ 
gauge symmetry involving
multilocal phase transformations of the wave functional which represents the
extended object. Thus, if $\Phi[\, \Sigma\, ]$ is the (complex) functional of 
the membrane configuration $\Sigma: X^\mu(\, \xi^1,\xi^2\, )$ modelling the 
bubble in terms of local coordinates $(\, \xi^1,\xi^2\, )$, then the action 
functional must be invariant under the combined transformation
$$
\delta_\Lambda\,  A_{\mu\nu\rho}=\partial_{[\,\mu}\, \Lambda_{\nu\rho\, ]}
\qquad
(\Lambda_{\nu\rho}=-\Lambda_{\rho\sigma})
\eqno(3)
$$
$$
\Phi[\, \Sigma\, ]\rightarrow \Phi'[\,\Sigma\, ]=
\exp\left[\, {i\over2}\, \int_{\Sigma=\partial\Omega}
dX^\mu\wedge dX^\nu\, \Lambda_{\mu\nu}(X)\, \right]\, \Phi[\Sigma]\ .
\eqno(4)
$$
This extended $U(1)$ gauge symmetry, if spontaneously broken, has the
same effect as the {\it global} $U(1)_{\rm PQ}$-symmetry as far as the
generation of Nambu-Goldstone bosons is concerned. However, one difference is 
that because of eq.(3), such Goldstone bosons will be realized in the 
Kalb-Ramond representation of massless spin-0 fields$^6$. A second and more
important difference is that such bosons can be gauged away leaving only
massive pseudoscalar particles in the physical spectrum. This is the essence
of DHM.
\medskip
The effect of the above properties on the inflationary-axion scenario is 
apparent in both the classical and quantum approach to Bubble Dynamics:

i) {\it Classical formulation.} The action functional of Bubble Dynamics
can be defined in any number of dimensions as a generalization of the
Einstein-Maxwell action for the dynamics of point charges on a Riemann
manifold$^7$. In four dimensions and under the assumption of spherical
symmetry, the field equations of bubble dynamics are integrable$^4$:
the net physical result of the $\a$-{\it bubble coupling} is the nucleation
of a bubble whose boundary separates two vacuum phases of De Sitter type
characterized by two effective and distinct cosmological constants, one 
inside and one outside the bubble. The evolution of the bubble,
which is controlled
by the two cosmological constants and by the surface tension, can be simulated
by the one-dimensional motion of a fictitious particle in a potential;
furthermore, a well defined algorithm exists$^{3,8}$ which is capable of 
determining all possible types of solutions, including inflationary ones,
together with the region in parameter space where simple families of solutions 
can exist.

ii) {\it Quantum mechanical formulation.} The cosmological field represents
the ultimate source of energy in the bubble universe. But how does matter
manage to `` bootstrap '' itself into existence out of that source of latent
energy? One possible scenario is as follows: as the volume of the bubble
increases exponentially during the inflationary phase, so does the total
(volume)energy of the De Sitter vacuum. At least classically. 
Quantum mechanically
there is a competitive effect which is best understood in terms of an
analogous effect in (1+1)-dimensions. In two dimensions a bubble is simply 
a particle-anti particle pair, moving `` left '' and `` right '' respectively,
and the volume within the bubble is the linear distance between them. As
the distance increases, so does the potential energy between them.
Quantum mechanically, however, it is energetically more favorable to
polarize the vacuum through the process of pair creation$^9$, which we
interpret as the nucleation of secondary bubbles out of the vacuum enclosed
by the original bubble. The net physical result of this mechanism is the
production of massive spin-0 particles$^{10}$. The same mechanism can be
lifted to (3+1)-dimensions and reinterpreted in the cosmological context:
the $\a$ field shares the same properties of the gauge potential $A_\mu$ in
two dimensions and polarizes the vacuum via the formation of secondary
bubbles$^{11}$.

Consider now a spherical bubble and focus on the radial evolution alone.
The intersection of any diameter with the bubble surface evolves precisely
as a particle-anti particle pair in (1+1)-dimensions. However, since there is
no preferred direction, the mechanism operates on concentric {\it shells} 
inside the original bubble. Remarkably, the final result is again the
production of massive pseudoscalar particles in the bubble universe.
However, while in two dimensions Goldstone bosons do not exist, in 
(3+1)-dimensions they do exist and have a direct bearing on the axion
mass problem.

A second possible scenario, which best illustrates how DHM works, is the
`` Membrane Abelian Higgs Model '', recently proposed by the authors$^{12}$.
For imaginary surface tension the absolute value of the membrane field can
acquire a non vanishing (constant) vacuum expectation value and the extended
gauge symmetry (3-4) is spontaneously broken; Goldstone bosons appear as
Kalb-Ramond fields $\Theta_{\mu\nu}$. By rotating the original field variables
to the unitary gauge, such Goldstone bosons are gauged away and one is left
with a residual lagrangian which leads precisely to the field equations (1).

While the mathematical details are far from trivial$^{12}$, the formal steps 
outlined above are familiar from the usual Higgs mechanism. The {\it dual} 
Higgs mechanism, however, differs in one crucial aspect. While in the usual 
Higgs mechanism the spin content of the gauge field is the same before and after
the appearance of mass, a new effect occurs when the gauge field is $\a$:
when massless, $\a$ is pure gauge, while equation (1) describes a massive 
particle with a definite value of spin (zero) in a representation which
is dual to the familiar Proca representation of massive, spin-1 particles:
the scalar component $\partial^\mu\, A^*_\mu$ of the pseudovector field 
$A^*_\mu$ is allowed to propagate as a free field by absorbing the Goldstone 
bosons.

It is a twist of Quantum Bubble Dynamics that the physical spectrum consists
of massive pseudo-scalar particles and one might be tempted to identify them
with the physical axions that might be present in our galactic halo. However,
the difference in DHM is plain: axions are massless spin-0 Goldstone bosons
represented by Kalb-Ramond fields while the physical spectrum consists of
massive spin-0 particles represented by the $\a$-field.

Born out of the darkness of the cosmic vacuum, axions were invisible to
begin with and remain invisible to the extent that they are `` eaten up ''
by the cosmological field. According to DHM, what experimenters are searching
for in the halo of our galaxy is the end product of a transmutation 
process which converts the self energy of the cosmological field into massive
particles. 
\vfill\eject

\centerline{\bf References}
\bigskip
\noindent
\item{1)} G. Efstathiou, W.J.Sutherland, S.J.Maddox, 
Nature,{\bf 348},(1990) 705 
\smallskip
\item{2)} V.A.Berezin, V.A.Kuzmin, I.I. Tkachev, Phys.Rev.{\bf D36}, (1987) 2919
\smallskip
\item{3)} S.Blau, G.Guendelman, A.Guth, Phys.Rev.{\bf D35}, (1987) 1474
\smallskip
\item{4)} A.Aurilia, G.Denardo, F.Legovini, E.Spallucci, Phys.Lett.{\bf B147},
(1984) 303; \hfill\break
A.Aurilia, G.Denardo, F.Legovini, E.Spallucci
Nucl.Phys.{\bf B252}, (1984) 523\hfill\break
A.Aurilia, R.Kissack, R.Mann, E.Spallucci, Phys.Rev.{\bf D35},(1987) 2961
\smallskip
\item{5)} Y.Nambu, Phys.Rep.{\bf 23C}, (1976), 250 
\smallskip
\item{6)} M.Kalb, P.Ramond, Phys.Rev.{\bf D9}, (1974) 2273
\smallskip
\item{7)} A.Aurilia, E.Spallucci, `` The Role of Extended Objects in Particle 
Physics Theory and Cosmology '', in \hfill\break
Proceedings of the Trieste Conference on
Super-Membranes and Physics in 2+1 Dimensions, Trieste 17-21 July 1989;
ed. M.J.Duff, C.N.Pope, E.Sezgin
\smallskip
\item{8)} A.Aurilia, M.Palmer, E.Spallucci, Phys.Rev.{\bf D40}, (1989) 2511
\smallskip
\item{9)} J.Kogut, L.Susskind, Phys.Rev.{\bf D11}, (1975) 3594
\smallskip 
\item{10)} J.Schwinger, Phys.Rev.{\bf 128}, (1962) 2425
\item{11)} J.D.Brown, C.Teitelboim, Phys.Lett.{\bf 195B}, (1987) 177\hfill
\break
A.Aurilia, E.Spallucci, Phys.Lett.{\bf B251}, (1990) 39 \hfill\break
A.Aurilia, R.Balbinot, E.Spallucci, 
Phys.Lett.{\bf B262}, (1991) 222
\smallskip
\item{12)} A.Aurilia, F.Legovini, E.Spallucci, 
 Phys.Lett.{\bf B264}, (1991) 69 
\bye